\documentstyle[emulateapj]{article}

\def\gsim{\;\rlap{\lower 2.5pt
 \hbox{$\sim$}}\raise 1.5pt\hbox{$>$}\;}
\def\lsim{\;\rlap{\lower 2.5pt
   \hbox{$\sim$}}\raise 1.5pt\hbox{$<$}\;}

\def\spose#1{\hbox to 0pt{#1\hss}}
\def\lta{\mathrel{\spose{\lower 3pt\hbox{$\mathchar''218$}}
     \raise 2.0pt\hbox{$\mathchar''13C$}}}
\def\gta{\mathrel{\spose{\lower 3pt\hbox{$\mathchar''218$}}
     \raise 2.0pt\hbox{$\mathchar''13E$}}}
\newcommand{\beq}{\begin{equation}}
\newcommand{\eeq}{\end{equation}}

\lefthead{HAIMAN \& HUI}
\righthead{QSO CLUSTERING}

\begin{document}
\title{Constraining the Lifetime of Quasars from their Spatial Clustering}
\author{Zolt\'an Haiman\altaffilmark{1}}
\affil{Princeton University Observatory, Princeton, NJ 08544, USA \\email: zoltan@astro.princeton.edu}
\and
\author{Lam Hui}
\affil{Institute for Advanced Study, Olden Lane, Princeton, NJ 08544, USA \\
email: lhui@ias.edu}

\altaffiltext{1}{Hubble Fellow}

\vspace{0.75\baselineskip}
\submitted{ApJ, in press (submitted on Feb. 8. 2000)}

\begin{abstract}

The lifetime of the luminous phase of quasars is constrained by current
observations to be $10^{6} \lsim t_Q \lsim 10^8$ years, but is otherwise
unkown.  We model the quasar luminosity function in detail in the optical and
X--ray bands using the Press--Schechter formalism, and show that the expected
clustering of quasars depends strongly on their assumed lifetime $t_Q$.
We quantify this dependence, and find that existing measurements of the
correlation length of quasars are consistent with the range $10^{6} \lsim t_Q
\lsim 10^8$. We then show that future measurements of the power spectrum of
quasars out to $z\sim 3$, from the 2dF or Sloan Digital Sky Survey, can
significantly improve this constraint, and in principle allow a precise
determination of $t_Q$.  We estimate the systematic errors introduced by
uncertainties in the modeling of the quasar-halo relationship, as well as by
the possible existence of obscured quasars.
\end{abstract}

\keywords{cosmology: theory -- cosmology: observation -- quasars: formation -- large scale structure}

\section{Introduction}
\label{section:introduction}

A long outstanding problem in cosmology is the synchronized evolution of the
quasar population over the redshift range $0\lsim z\lsim 5$.  Observations in
the optical (Pei 1995) and radio (Shaver et al. 1994) show a pronounced peak in
the abundance of bright quasars at $z\approx 2.5$; recent X--ray observations
(Miyaji et al. 2000) confirm the rapid rise from $z=0$ towards $z\approx 2$,
but have not shown evidence for a decline at still higher redshifts.
Individual quasars are widely understood to consist of supermassive black holes
(BHs) powered by accretion (Lynden-Bell 1967; Rees 1984).  A plausible
timescale for quasar activity is then the Eddington time, $4\times10^7$
($\epsilon$/0.1) yr, the e-folding time for the growth of a BH accreting mass
at a rate $\dot M$, while shining at the Eddington luminosity with a radiative
efficiency of $L=L_{\rm Edd}=\epsilon\dot M c^2$.  The lifetime $t_Q$ of the
luminous phase of quasars can be estimated directly, by considering the space
density of quasars and galaxies.  At $z\sim 2$, the ratio $n_Q/n_G \sim 3\times
10^{-3}$ implies the reassuringly close value of $t_Q\sim t_{\rm Hub}
n_Q/n_G\sim 10^7$ yr (Blandford 1999 and references therein).  These lifetimes
are significantly shorter than the Hubble time, suggesting that the quasar
population evolves on cosmic time--scales by some mechanism other than local
accretion physics near the BH.

It is tempting to identify quasars with halos condensing in a cold dark matter
(CDM) dominated universe, as the halo population naturally evolves on cosmic
time--scales (Efstathiou \& Rees 1988; Haiman \& Loeb 1998; Kauffmann \&
Haehnelt 2000).  Furthermore, quasars reside in a subset of all galaxies, while
the redshift--evolution of the galaxy population as a whole (qualitatively
similar to that of bright quasars) has been successfully described by
associating galaxies with dark halos (e.g. Lacey \& Cole 1993; Kauffmann \&
White 1993).  A further link between galaxies and quasars comes from the recent
detection, and measurements of the masses of massive BHs at the centers of
nearby galaxies (Magorrian et al. 1998; van der Marel 1999).

These arguments suggest that the evolution of the quasar population can indeed
be described by ``semi--analytic'' models, associating quasars with dark matter
halos. In this type of modeling, the quasar lifetime plays an important role.
The quasar phase in a single halo could last longer ($t_{\rm Q}\sim10^8$yr),
with correspondingly small $M_{\rm bh}/M_{\rm halo}$ ratios, or last shorter
($t_{\rm Q}\sim10^6$yr), with larger BH formation efficiencies (Haiman \& Loeb
1998; Haehnelt et al. 1998).  Note that although recent studies have
established a correlation between the bulge mass $M_{\rm bulge}$ and BH mass
$M_{\rm bh}$, this correlation leaves a considerable uncertainty in the
relation between $M_{\rm bh}$ and the mass $M_{\rm halo}$ of its host halo.  If
the initial density fluctuations are Gaussian with a CDM power spectrum, the
clustering of collapsed halos is a function of their mass -- rarer, more
massive halos cluster more strongly (Kaiser 1984; Mo \& White 1996).  Hence,
measurements of quasar clustering are a potentially useful probe of both BH
formation efficiencies and quasar lifetimes (La Franca et al. 1998, Haehnelt et
al. 1998).

In this paper, we assess the feasibility of breaking the above degeneracy, and
inferring quasar lifetimes, from the statistics of clustering that will be
available from the Sloan Digital Sky Survey (SDSS, Gunn \& Weinberg 1995) and
Anglo-Australian Telescope Two-Degree-Field (2dF, Boyle et al. 1999).  Previous
works (e.g. Stephens et al.  1997; Sabbey et al. 1999) have yielded estimates
suggesting that quasars are clustered more strongly than galaxies. However, the
current uncertainties are large, especially at higher redshifts, where
clustering has been found to decrease (Iovino \& Shaver 1988; Iovino et
al. 1991), to stay constant (Andreani \& Cristiani 1992; Croom \& Shanks 1996),
or to increase with redshift (La Franca et al. 1998). As a result, no strong
constraints on the life--time can be obtained yet. The key advance of
forthcoming surveys over previous efforts is two--fold.  Because of their sheer
size, i.e. the large number of quasars covering a large fraction of the sky,
both shot--noise and sample variance can be beaten down, significantly reducing
the statistical uncertainties.  Furthermore, the large sample-size will
eliminate the need to combine data from different surveys with different
selection criteria, hence allowing cleaner interpretation.

Recent measurements of the local massive black hole density have stimulated
discussions of a radiative efficiency which is much lower than the usual $\sim
0.1$ (e.g. Haehnelt et al. 1999). A convincing constraint on the lifetime of
quasars could be therefore highly interesting, as this might have implications
for the local accretion physics near the BH.

This paper is organized as follows. In \S~\ref{section:models}, we summarize
our models for the quasar luminosity function, and in
\S~\ref{section:clustering} we compute the quasar correlation function in these
models.  In \S~\ref{section:present}, we compare the model predictions with
presently available data, and in \S~\ref{section:future} we assess the ability
of future optical redshift surveys to discriminate between the various models.
In \S~\ref{section:xray}, we repeat our analysis in the soft X--ray band, and
examine the contribution of quasars to the X-ray background, and its
auto--correlation.  In \S~\ref{section:discussion}, we address some of the
caveats arising from our assumptions, and in \S~\ref{section:conclusions} we
summarize our conclusions and the implications of this work.

\section{Models for the Quasar Luminosity Function}
\label{section:models}

In this section, we briefly summarize our model for the luminosity function
(LF) of quasars, based on associating quasar BHs with dark halos.  Our
treatment is similar to previous works (Haiman \& Loeb 1998; Haehnelt et
al. 1998), although differs in some of the details.  A more extensive treatment
is provided in the Appendix.  The main assumption is that there is, on average,
a direct monotonic relation between halo mass $M_{\rm halo}$ and average quasar
luminosity $L_{M,z}$, which we parameterize using the simple
power--law ansatz:

\beq \bar L_{M,z}= x_0(z) M_{\rm halo} \left(\frac{M_{\rm
halo}}{M_0}\right)^{\alpha(z)}.
\label{eq:params00}
\eeq

Here $x_0(z)$ and $\alpha(z)$ are ``free functions'', whose values are found by
the requirement that the resulting luminosity function agrees with
observations.  As explained in the Appendix, our model has one free parameter,
the lifetime $t_Q$, which uniquely determines $x_0(z)$ and $\alpha(z)$ in any
given background cosmology.  We assume the background cosmology to be either
flat ($\Lambda$CDM) with $(\Omega_\Lambda,\Omega_{\rm m},h,\sigma_{\rm
8h^{-1}},n)=(0.7,0.3,0.65,1.0,1.0)$ or open (OCDM) with
$(\Omega_\Lambda,\Omega_{\rm m},h,\sigma_{\rm
8h^{-1}},n)=(0,0.3,0.65,0.82,1.3)$.  In LCDM, we find $(-\log[x_0/{\rm L_\odot
M_\odot^{-1}}],\alpha) \approx (1,0.4)$ and $\approx (-0.2, -0.1)$ for
lifetimes of $t_Q=10^{8}$ and $t_Q=10^{6.5}$yr, respectively. Similarly, in
OCDM, we find $(-\log[x_0/{\rm L_\odot M_\odot^{-1}}],\alpha) \approx (1,0.25)$
and $\approx (-0.2, -0.25)$ for these two lifetimes.

In Figure~\ref{fig:LFfits}, we demonstrate the agreement between the LF
computed in our models with the observational data at two different redshifts,
$z=2$ and $z=3$.  For reference, the upper labels in this figure show the
apparent magnitudes in the SDSS $g^\prime$ band, assuming that the intrinsic
quasar spectrum is the same as the mean spectrum in the Elvis et al. (1994)
quasar sample. The photometric detection threshold\footnote{See
http://www.sdss.org/science/tech\_summary.html.} of SDSS is $g^\prime \approx
22.6$, corresponding to a BH mass of $\approx 10^{8} {\rm M_\odot}$ at $z=3$
and a three times smaller mass at $z=2$. As the figure shows, the overall
quality of the fits is excellent; for reference, the dashed lines show the
ad--hoc empirical fitting formulae from Pei (1995).  Similarly accurate match
to the quasar LF is achieved at different redshifts, and in the models assuming
an OCDM cosmology.  Figure~\ref{fig:LFfits} shows, in particular, that the fits
obtained from the power--law ansatz adopting either a short (solid lines) or a
long (dotted lines) lifetime are nearly indistinguishable; hence modeling the
LF by itself does not constrain the quasar lifetime within the limits
$10^{6.5}$ yr $\lsim t_Q \lsim 10^8$ yr.

Before considering constraints on the lifetime from clustering, it is useful to
point out that estimates for both upper and lower limits on $t_Q$ follow from
the observed luminosity function alone.

{\it Lower limit on $t_Q$.} A halo of mass $M_{\rm halo}$ is unlikely to harbor
a BH more massive than $\approx 6\times10^{-3} (\Omega_{b}/\Omega_{0}) M_{\rm
halo} = 6\times 10^{-4} M_{\rm halo}$, where $\approx 6\times10^{-3}$ is the
ratio $M_{\rm bh}/M_{\rm bulge}$ found in nearby galaxies (Magorrian et
al. 1998) because $M_{\rm bulge}$ cannot be larger than
$(\Omega_{b}/\Omega_{0}) M_{\rm halo}$.  This maximal BH could at best emit
$\approx 10\%$ of the Eddington luminosity in the B--band, implying $L/M_{\rm
halo} \lsim 3~{\rm L_\odot/M_\odot}$.  We find (cf. Fig.~\ref{fig:params}) that
our short lifetime model with $t_Q=10^{6.5}$ yr nearly reaches this limit;
models with shorter lifetimes would require unrealistically large $L/M_{\rm
halo}$ ratios.

{\it Upper limit on $t_Q$.} Long lifetimes, on the other hand, require the
ratio $L/M_{\rm halo}$ to be small; this can lead to unrealistically large halo
masses.  The brightest quasars detected at redshifts $z\approx 2-3$ have
luminosities as large as $L\approx 10^{14} {\rm L_\odot}$ (Pei 1995). We find
(cf. Fig.~\ref{fig:params}) that in order to avoid the host halo masses of
these bright quasars to exceed $\sim 10^{15}~{\rm M_\odot}$, the lifetime
cannot be longer than $\sim 10^8$ yr. An alternative, standard argument goes as
follows.  The black hole mass grows during the quasar phase as $e^{t_Q/t_E}$
where $t_E = 4\times10^7 (\epsilon/0.1)$ yr is the Eddington time. Assuming a
conservative initial black hole mass of $1 M_{\odot}$, a lifetime longer than
$10^9$ yr would give final black hole masses that are unacceptably large.

\vspace{2\baselineskip}

\section{The Clustering of Quasars}
\label{section:clustering}

As demonstrated in the previous section, equally good fits can be obtained to
the luminosity function of quasars, assuming either a short or a long lifetime,
and a power-law dependence of the mean quasar luminosity on the halo mass $\bar
L\propto M^\alpha$.  In this section, we derive the clustering of quasars in
our models, and demonstrate that they depend significantly on the assumed
lifetime.

The halos are a biased tracer of the underlying mass distribution, customarily
expressed by $P_{\rm halo} (k) = b^2 P(k)$ where $P_{\rm halo}$ and $P$ are the
halo and mass power spectra as a function of wavenumber $k$. The bias parameter
for halos of a given mass $M$ at a given redshift $z$ is given by (Mo \& White
1996)

\beq
b(M,z)= 1 +  \frac{1}{\delta_c}\left[
\left(\frac{\delta_c}{\sigma(M)D(z)}\right)^2-1\right],
\label{eq:biasM}
\eeq

where $D(z)$ is the linear growth function, $\sigma(M)$ is the r.m.s. mass
fluctuation on mass--scale $M$ (using the power spectrum of Eisenstein \& Hu
1999), and $\delta_c\approx 1.68$ is the usual critical overdensity in the
Press--Schechter formalism (see Jing 1999 and Sheth \& Tormen 1999 for more
accurate expressions for $b(M,z)$ for low $M$, which we find not to affect our
results here).  The bias associated with quasars with luminosity $L$ in our
models is given by averaging over halos of different masses associated with
this luminosity. Following equation \ref{eq:matchdiff}, we obtain

\begin{eqnarray}
b(L,z) = && \hspace{-0.5cm}\left[\frac{d\phi}{dL}(L,z)\right]^{-1} \times
\int_0^\infty dM  \frac{dN}{dM}(M,z) \\
\nonumber && b(M,z) \frac{dg}{dL}(L,\bar L_{M,z}) f_{\rm on}(M,z).
\label{eq:biasL}
\end{eqnarray}

We show in Figure~\ref{fig:bias} the resulting bias parameter $b(L,z)$ in the
models corresponding to Figure~\ref{fig:LFfits}, with short and long lifetimes,
and at redshifts $z=2$ and 3.  As expected, quasars are more highly biased in
the long lifetime model, by a ratio $b$(long)/$b$(short)$\gsim 2$.  In the
$\Lambda$CDM case, at the detection threshold of SDSS, we find
$b$(long)$\approx 3$ at $z=3$ and $b$(long)$\approx2$ at $z=2$. Bright quasars
with $g^\prime=17$ are predicted to have a bias at $z=3$ as large as $b=10$.
For reference, we also show in this figure the bias parameters obtained in the
OCDM cosmology, which are significantly lower than in the $\Lambda$CDM case.
The number of quasars observed at a fixed flux implies an intrinsically larger
number of sources if OCDM is assumed, because the volume per unit redshift and
solid angle in an open universe is smaller.  This lowers the corresponding halo
mass and therefore the bias.

\section{Comparison with Available Data}
\label{section:present}

As emphasized in \S~\ref{section:introduction}, the presently available data
leave considerable uncertainties in the clustering of quasars. Nevertheless, it
is interesting to contrast the results of the previous section with preliminary
results from the already relatively large, homogeneous sample of high--redshift
quasars in the 2dF survey (Boyle et al. 2000). Our predictions are obtained by
relating the apparent magnitude limit to a minimum absolute luminosity at a
given redshift: $\log [L_{\rm min}(z)/L_{B,\odot}] = 0.4[5.48- B + 5\log(d_{\rm
L}(z)/{\rm pc})-5]$.  This relation assumes no K--correction, justified by the
nearly flat quasar spectra ($\nu F_\nu = $ const) at the relevant wavelengths
(e.g. Elvis et al. 1994; Pei 1995). In our model, the correlation length $r_0$
is given implicitly by

\beq
\xi_q(r_0)\equiv \bar b^2(z) D^2(z) \xi_m(r_0) = 1,
\label{eq:xiq}
\eeq

where $\xi_m(r)$ is the usual dark matter correlation function, and $\bar b(z)$
is the value of the bias parameter $b(L,z)$ as determined in the previous
section, but now averaged over all quasars with magnitudes brighter than the
detection limit,

\beq
\bar b(z) = \left[\int_{L_{\rm min}(z)}^\infty dL \frac{d\phi}{dL} \right]^{-1}
\int_{L_{\rm min}(z)}^\infty dL \frac{d\phi}{dL} b(L,z).
\label{eq:bave}
\eeq

In Figure~\ref{fig:r0} we show the resulting correlation lengths in the long
and short lifetime models.  Also shown is a preliminary data--point with
1$\sigma$ error--bars from the 2dF survey, based on $\approx$3000 quasars with
apparent magnitudes $B < 20.85$ (Croom et al. 1999).  The upper panel shows the
results in our fiducial $\Lambda$CDM model with predictions for this magnitude
cut.  The published results for $r_0$ are cosmology dependent, and we have
simply converted them for our cosmological models by taking the corresponding
average of the redshift-distance and angular-diameter-distance -- this crude
treatment is adequate given the large measurement errors.  Our models
generically predict a gradual increase of the correlation length with redshift
(``positive evolution''). The clustering is dominated by the faintest quasars
near the threshold luminosity; as a result, the fixed magnitude-cut of
$B=20.85$ corresponds to more massive, and more highly clustered halos at
higher redshifts.  There are two additional effects that determine the
redshift--evolution of clustering: (1) quasars of a fixed luminosity are more
abundant towards high--$z$, requiring smaller halo masses to match their number
density, and (2) halos with a fixed mass are more highly clustered towards
high-$z$.  We find, however, that these effects are less important than the
increase in $M_{\rm halo}$ caused by fixing the apparent magnitude threshold,
which gives rise to the overall positive evolution.

The clustering in the long--lifetime model is stronger, and evolves more
rapidly than in the short--lifetime case.  As we can see, the present
measurement error-bars are large -- the whole range of life-times from
$10^{6.5}$ to $10^8$ yr is broadly consistent with the data, to within $\sim 2
\sigma$.  For the $\Lambda$CDM model, the 2dF data-point is consistent with a
lifetime of around $10^{7.7 \pm 0.8}$ yr; in the OCDM model the lifetime is
somewhat higher, $10^{8 \pm 0.8}$ yr. It is worth emphasizing here that the
constraints on lifetime from clustering depends on the underlying cosmological
parameters one assumes, which future large scale structure measurements (from
e.g. the microwave background, galaxy surveys and the Lyman--$\alpha$ forest)
will hopefully pin down to the accuracy required here.

Lastly, we note that observational results on $r_0$ are commonly obtained by a
fit to the two-point correlation of the form $\xi (r) =
(r/r_0)^{-\gamma}$. Since $r_0$ is the correlation length where the correlation
is unity, we expect our formalism to begin to fail on such a scale, because
neither linear fluctuation growth nor linear biasing holds. On the other hand,
it is also unclear whether $r_0$ as presently measured from a 2-parameter fit
to the still rather noisy observed two-point correlation necessarily
corresponds to the true correlation length.  While our crude comparison with
existing data in Figure~\ref{fig:r0} suffices given the large measurement
errors, superior data in the near future demand a more refined treatment, which
is the subject of the next section.

\section{Expectations from the SDSS and 2dF}
\label{section:future}

Although existing quasar clustering measurements still allow a wide range of
quasar lifetimes, and do not provide tight constraints on our models,
forthcoming large quasar samples from SDSS or the complete 2dF survey are
ideally suited for this purpose. Here we estimate the statistical uncertainties
on the derived lifetimes, using the three--dimensional quasar power spectrum
$P_Q(k)$ derived from these surveys.

The variance of the power spectrum is computed by

\beq 
\langle \delta P_{\rm Q}^2 (k) \rangle = n_k^{-1} [\bar b^2 P(k) + \bar n]^2,
\label{eq:variance}
\eeq 

where the large scale fluctuations are assumed to be Gaussian, $n_k$ is the
number of independent modes, $\bar n$ is the mean number density of observed
quasars, $P_Q(k) = \bar b^2 P(k)$ is the quasar power spectrum and $P(k)$ is
the mass power spectrum (Feldman, Kaiser \& Peacock 1994).  For a survey of
volume $V$, and a $k$-bin of size $\Delta k$, we use $n_k = k^2 \Delta k V / 4
\pi^2$.  The fractional variance is therefore $n_k^{-1} \{1 + 1/[\bar n \bar b
P(k)]\}^2$. In terms of minimizing this error, increasing the luminosity cut of
a survey has the advantage of raising the bias $\bar b$, but has the
disadvantage of decreasing the abundance $\bar n$. In practice, $\bar b$
changes relatively slowly with mass (slower than $\bar b\propto M$) whereas
$\bar n$ varies with mass much more rapidly ($\bar n \sim 1/M$, or steeper if
$M>M_\star$).  As a result, we find that for our purpose of determining the
clustering and the quasar lifetime, it is better to include more (fainter)
quasars.

The power spectrum $P_Q(k)$ of quasars in $\Lambda$CDM is shown in
Figure~\ref{fig:sdsslcdm} at two different redshifts near the peak of the
comoving quasar abundance, $z=2$ and $z=3$.  We assume that redshift slices are
taken centered at each redshift with a width of $\Delta z = 0.5$ (which enters
into the volume $V$ above).  Results are shown in the long and short lifetime
models, together with the expected $1\sigma$ error bars from SDSS (crosses).
Also shown in the lower panel are the expected error bars from 2dF (open
squares), which are slightly larger because of the smaller volume (for SDSS, we
assume an angular coverage of $\pi$ steradians, and for 2dF, an area of $0.23$
steradians). We do not show error bars for 2dF beyond $k \sim 0.01 {\, \rm
Mpc/h}$ because larger scales would likely be affected by the survey window.
We also only show scales where the mass power spectrum and biasing are believed
to be linear.

As these figures show, the long and short $t_Q$ models are easily
distinguishable with the expected uncertainties both from the SDSS and the 2dF
data, out to a scale of $\sim 100$Mpc.  The measurement errors at different
scales are independent (under the Gaussian assumption), and hence, when
combined, give powerful constraints e.g. formally, even models with lifetimes
differing by a few percent can be distinguished with high confidence using the
SDSS.  However, systematic errors due to the theoretical modeling are expected
to be important at this level, which we will discuss in \S
\ref{section:discussion}.

In Figure~\ref{fig:sdsslcdm}, we have used the magnitude cuts for spectroscopy,
i.e. $B=20.85$ for 2dF and $B=20.4$ ($g^\prime\approx 19$) for SDSS.  If
photometric redshifts of quasars are sufficiently accurate, the magnitude cuts
can be pushed fainter, further decreasing the error-bars -- although it is
likely that the photometric redshift errors will be large enough that one can
only measure the clustering in projection, in some well-defined redshift-bin
picked out using color information. Color selection of $z > 3$ quasars has
already proven to be highly effective (Fan et al. 1999); a high redshift sample
can give valuable information on the evolution of the quasar clustering (see
Fig. \ref{fig:r0}).

\section{Quasar Clustering in X-ray}
\label{section:xray}

Both the luminosity function (e.g. Miyaji et al. 2000), and the clustering of
quasars (e.g. Carrera et al. 1998) has been studied in the X--ray band,
analogously to the optical observations described above. At present, the
accuracy of both quantities are inferior to that in the optical.  Nevertheless,
it is interesting to consider the clustering of quasars in the X--ray regime,
because (1) applying the exercise outlined above to a different wavelength band
provides a useful consistency check on our results, (2) observations with the
{\it Chandra X--ray Observatory (CXO)} and {\it XMM} can potentially\footnote{
see http://asc.harvard.edu and http://xmm.vilspa.esa.es, respectively} probe
quasars at redshifts higher than currently reached in the optical, and (3)
X-ray observations are free from complications due to dust extinction, although
other forms of obscuration are possible (see \S \ref {section:discussion}). In
addition, we will consider here the auto--correlation of the soft X--ray
background (XRB) as another potential probe of quasar clustering and lifetime.

The formalism we presented in \S~\ref{section:models} and
\S~\ref{section:clustering} is quite general, and we here apply it to the soft
X--ray luminosity function (XRLF) from Miyaji et al. (2000).  The details of
the fitting procedure are given in the Appendix. In analogy with the optical
case, we find that the clustering of X-ray selected quasars depends strongly on
the lifetime.  As an example, including all quasars whose observed flux is
above $3 \times 10^{-14} {\, \rm erg \, cm^{-2}\, s^{-1}}$, we find the
correlation length at $z=2$ to be $\approx 4h^{-1}$Mpc in the short lifetime,
and $\approx 11h^{-1}$Mpc in the long lifetime case ($\Lambda$CDM).  Current
data probes the clustering of X-ray quasars only at low redshifts ($z\lsim 1$,
see Carrera et al. 1998), where our models suffer from significant
uncertainties due to the subhalo problem discussed in \S
\ref{section:discussion}. Constraints at $z\gsim 2$ could be available in the
future from {\it CXO} and {\it XMM}, provided that a large area of the sky is
surveyed at the improved sensitivities of these instruments.

We next focus on the quasar contribution to the soft X--ray background and its
auto--correlation, which as we will see is dominated by quasar contributions at
somewhat higher redshifts. The mean comoving emissivity at energy $E$ from all
quasars at redshift $z$, typically in units of ${\rm keV \, cm^{-3} \, s^{-1}
\, sr^{-1}}$, is given in our models by

\beq
\bar j(E,z) = \frac{1}{4\pi} \int_0^\infty dL \frac{d\phi}{dL} L_X(E,L),
\label{eq:jxray}
\eeq where $L_X(E,L)$ is the luminosity (in ${\rm keV \, s^{-1} \, keV^{-1}}$)
at the energy $E$ of a quasar whose luminosity at $(1+z)$ keV is $L$.  We have
used the mean spectrum of Elvis et al. (1994) to include a small K--correction
when computing the background at observed energies $E\ne 1$keV.  The mean
background is the integral of the emissivity over redshift,

\begin{equation}
\bar I(E)  =  \int_0^\infty \frac{d\chi}{(1+z)} \bar j(E_z,\chi),
\label{eq:xrb}
\end{equation}
where $\bar I$ is typically given in units of ${\rm keV \, cm^{-2} \, s^{-1} \,
sr^{-1} \, keV^{-1}}$, $E_z = E (1+z)$, and $\chi$ is the comoving distance
along the line of sight.

If $\delta(z)$ is the mass fluctuation at some position at redshift $z$, then
the fluctuation of the emissivity at the same position and redshift is given by
$b_X(z) \delta(z) \bar j(E_z, z)$, where we have defined the X-ray
emission--weighted bias $\bar b_X(z)$ as

\beq
\bar b_X(z) = \frac{1}{\bar j(E_z,z)} 
\frac{1}{4\pi} \int_0^\infty dL \frac{d\phi}{dL} L_X(E_z,L) b_X(L,z).
\label{eq:biasx}
\eeq

For simplicity, we compute the auto--correlation $w_\theta$ of the XRB using
the Limber approximation, together with the small angle approximation, as:

\begin{eqnarray}
\label{eq:Clwtheta}
C_\ell(E) &=& {\bar I(E)}^{-2} \int \frac{d\chi}{r_\chi^2}
W^2(E_\chi,\chi) P_0(\ell/r_\chi) \\ 
\nonumber w_\theta(E) &=& 
\int {\ell d\ell \over 2\pi} C_\ell(E) J_0 (\ell \theta),
\end{eqnarray}

where $C_\ell$ is the angular power spectrum, $J_0$ is the zeroth order Bessel
function, $P_0(\ell/r_\chi)$ is the linear mass power spectrum today, $r_\chi$
is the angular diameter distance ($=\chi$ for a flat universe), and
$W(E_\chi,\chi) = \bar j(E_z,z) \bar b_X(z) D(z)/(1+z)$. When the power
spectrum is measured in practice, shot-noise has to be subtracted or should be
included in the theoretical prediction, whereas the same is not necessary for
the angular correlation except at zero-lag.

Our models predicts the correct mean background spectrum $\bar I(E)$, computed
from equation~\ref{eq:xrb}, at $E=1$ keV.  We have included all quasars down to
the observed 1 keV flux of $2\times10^{-17}~{\rm erg~cm^{-2}~s^{-1}}$, i.e. we
used our models to extrapolate the XRLF to two orders of magnitude fainter than
the ROSAT detection threshold for discrete sources (Hasinger \& Zamorani 1997),
to make up the remaining $\sim 50\%$ of the XRB at 1keV. Our models predict a
faint--end slope that is steeper than the Miyaji et al. (2000) fitting
formulae, allowing faint quasars to contribute half of the background.  The
emissivities peak at $z\approx 2$, coinciding with the peak of the XRLF,
implying that our model produces most of the XRB, as well as its
auto--correlation signal at $z\approx 2$.  Note that the known contribution
from nearby galaxy clusters is $\sim 10\%$ (Gilli et al. 1999), which we ignore
here.

In Figure \ref{fig:xrb}, we show our predictions for the two point angular
correlation $w_\theta$ of the XRB from quasars at 1 keV. Most measurements at
the soft X-ray bands have yielded only upper limits, which are consistent with
our predictions (e.g.  variance at $\lsim 0.12$ at a scale of $10$ arcmin.  and
$E = 0.9 - 2$keV, from Carrera et al. 1998; see also references therein).
Soltan et al. (1999) obtained angular correlations significantly higher than
previous results (dashed curve), which, taken at face value, would imply quasar
lifetimes $t_Q \gg 10^8$ yr.  However, the results of Soltan et al. (1999)
could be partially explained by galactic contamination (Barcons et
al. 2000). We therefore view this measurement as an upper limit, which is
consistent with models using both lifetimes we considered. Figure \ref{fig:xrb}
shows that $w_\theta$ predicted in the long and short lifetime models differ by
a factor of $\sim 2$ on angular scales of 0.1-1 degrees, offering another
potential probe of quasar lifetimes, provided that $w_\theta$ can be measured
more accurately in the future, and that the contribution to the clustering
signal from nearby non-quasar sources (e.g. clusters) is small or can be
subtracted out.  Finally, we note that there have been detections of of
clustering on several-degree-scales at the hard X-ray bands ($2 - 10$ keV) from
the HEAO satellite (Treyer et al. 1998) -- while a prediction for such energies
would be interesting (see also Lahav et al. 1997), it would require an
extrapolation of the X-ray spectrum, since we normalize by fitting to the
soft-Xray luminosity function.

\section{Further Considerations}
\label{section:discussion}

We have shown above that the quasar lifetime could be measured to high
precision, using the soon available large samples of quasars at $z\lsim 3$;
either from the 2dF or the SDSS survey.  This precision, however, reflects only
the statistical errors in the simple model we have adopted for relating quasars
to dark matter halos. The main hindrance in determining the quasar lifetime
will likely be systematic errors; here we discuss how several potential
complications could affect the derived lifetime.

\vspace{\baselineskip} {\em Obscured sources.} Considerations of the hard
X--ray background have led several authors to suggest the presence of a large
population of ``absorbed'' quasars, necessary to fit the hard slope and overall
amplitude of the background.  Although not a unique explanation for the XRB,
this would imply that the true number of quasars near the faint end of both the
optical and soft X--ray LF is $\sim10$ times larger than what is observed; 90\%
being undetected due to large absorbing columns of dust in the optical, and
neutral hydrogen in the soft X--rays (see, e.g. Gilli et al. 1999).  Unless the
optically bright and dust--obscured phases occur within the same object (Fabian
\& Iwasawa 1999), this increase would have a direct effect on our results,
since we would then need to adjust our fitting parameters to match a $\sim10$
times higher quasar abundance.  We find that this is easily achieved by leaving
$x_0$ and $\alpha$ unchanged, and instead raising the lifetime from $10^{6.5}$
to $10^{7.5}$ yr in the short lifetime model, and from $10^{8}$ to $10^{8.6}$
yr in the long lifetime model.  In the latter case, a 10-fold increase in the
duty cycle requires only an increase in $t_Q$ by a factor of $\approx 4$, owing
to the shape of the age distribution $dp_a/dt$ (Lacey \& Cole 1993).  Our
results would then hold as before, but they would describe the two cases of
$t_Q=10^{7.5}$ and $10^{8.6}$ yr.  Interestingly, this scenario would imply
that the quasar lifetime could not be shorter than $t_Q\approx10^{7.5}$ yr,
simply based on the abundance of quasars (cf. \S~\ref{section:models}).  Future
infrared and hard X-ray observations should help constrain the abundance of
obscured sources, and reduce this systematic uncertainty.

\vspace{\baselineskip} {\em Multiple BH's in a single halo.}  A possibility
that could modify the simple picture adopted above is that a single halo might
host several quasar black holes.  A massive (e.g. $10^{14}~{\rm M_\odot}$) halo
corresponds to a cluster of galaxies; while the Press--Schechter formalism
counts this halo as a single object.  If quasar activity is triggered by
galaxy--galaxy mergers, a massive Press-Schechter halo, known to contain
several galaxies, could equally well host several quasars (e.g. Cavaliere \&
Vittorini 1998). There is some observational evidence of perhaps merger driven
double quasar activity (Owen et al. 1985; Comins \& Owen 1991). One could
therefore envision that quasars reside in the sub--halos of massive ``parent''
halos -- a scenario that would modify the predicted clustering. To address
these issues in detail, one needs to know the mass--function of sub--halos
within a given parent halo, as well as the rate at which they merge and turn
on.  In principle, this information can be extracted from Monte Carlo
realizations of the formation history of halos in the extended Press--Schechter
formalism (i.e. the so called merger tree) together with some estimate of the
time-scale for mergers of sub--halos based on, for instance, dynamical friction
(e.g. Kauffmann \&  Haehnelt 2000).  Here, we consider two toy models that we
hope can bracket the plausible range of clustering predictions.

To simplify matters, we ignore the scatter in $L$--$M$ in the following
discussion.  Suppose one is interested in quasars of a luminosity $L$ at
redshift $z_0$, which correspond to Press-Schechter halos of mass $M_0$ in our
formalism as laid out in \S \ref{section:models}. This choice of $M_0$ matches
the abundance of quasars, expressed approximately as $L(d\Phi/dL)\approx
M_0[dN(M_0, z_0)/dM_0] (t_Q/t_{\rm Hub})$ (the merger, or activation rate of
halos is approximated as $\sim t_{\rm Hub}^{-1}$, where $t_{\rm Hub}$ is the
Hubble time), and implies the bias $b_0(L)\approx b(M_0,z_0)$.

In model A, we suppose the Press-Schechter halos are identified at some earlier
redshift $z_1$ -- these would be sub--halos of those Press-Schechter halos
identified at $z_0$. Quasars of luminosity $L$ now correspond to sub--halos of
mass $M_1$.  The abundance of these sub-halos is given by the Press-Schechter
mass function $dN(M_1, z_1)/dM_1$, which is related to the mass function at
$z_0$ by $dN(M_1, z_1)/dM_1 = \int_{M_1}^\infty dM [dN(M, z_0)/dM] [d{\cal
N}(M_1, z_1 | M, z_0) / dM_1]$ where $dM_1 \times d{\cal N}(M_1, z_1 | M, z_0)
/ dM_1$ is the average number of $M_1 \pm dM_1/2$ sub--halos within parent
halos of mass $M$, given by (e.g. Sheth \& Lemson 1999)

\begin{eqnarray}
\label{eq:condPS}
&& d{\cal N}(M_1, z_1 | M, z_0) / dM_1 = \\ \nonumber
&& \quad \quad {M\over M_1} {1\over \sqrt{2\pi}} 
{{\delta_1 - \delta_0}\over {[\sigma^2 (M_1) - \sigma^2 (M)]^{3/2}}} \\ \nonumber
&& \quad \quad
{\, \rm exp} \left\{ -{(\delta_1 - \delta_0)^2 \over {2[\sigma^2 (M_1)-\sigma^2 (M)]}}\right\}
{d \sigma^2 (M_1) \over dM_1},
\end{eqnarray}

where $\delta_1 = \delta_c/D(z_1)$ and $\delta_0 = \delta_c/D(z_0)$ (see
eq. \ref{eq:biasM}).

To match the abundance of quasars at luminosity $L$, we impose the condition
that $M_1 [dN(M_1, z_1)/dM_1] \approx$ $M_0[dN(M_0, z_0)/dM_0]$ , which
determines $M_1$ given $z_1$, $M_0$ and $z_0$. The bias of the quasars is no
longer $b_0 (L) \approx b(M_0, z_0)$, but is instead given by

\begin{eqnarray}
\label{eq:beff}
&& b_{\rm eff}^A (L,z_1) = \left[{dN(M_1, z_1) \over dM_1}\right]^{-1} \\ \nonumber 
&& \quad \int_{M_1}^\infty dM {dN(M, z_0) \over dM} {d{\cal N}
(M_1, z_1 | M, z_0) \over dM_1} b(M, z_0).
\end{eqnarray}

We show in Fig. \ref{fig:comparebiasPS} the ratio of $b_{\rm eff}^A (L,z_1)/b_0
(L)$ as a function of $z_1$, for $z_0 = 3$ and $z_0 = 2$ respectively, and for
a range of masses $M_0$ which are representative of the halos that dominate our
clustering signal in previous discussions.  It is interesting how the bias
$b_{\rm eff}^A (L,z_1)$ is not necessarily larger than our original bias $b_0
(L)$, despite the fact that the bias of sub--halos should be boosted by their
taking residence in bigger halos.  This is because the relevant masses here
(e.g. $M_0$) are generally large, and we find that the number of halos of mass
$M_0$ at $z_1$ is always {\it smaller} than the number of halos of the same
mass at $z_0 < z_1$. Hence, to match the observed abundance of quasars at the
same $L$, $M_1$ must be chosen to be smaller than $M_0$.  As
Fig. \ref{fig:comparebiasPS} shows, this could, in some cases, more than
compensate the increase in clustering due to massive parent halos.  Because of
these two opposing effects, the bias does not change by more than about $50 \%$
even if one considers $z_1$ as high as $10$.  This translates into a factor of
$\sim 2$ uncertainty in our predictions for the quasar power spectrum. Our
clustering predictions for the short and long lifetime models differ by about a
factor of $\sim 5$, implying that the lifetime can still be usefully
constrained at $z\gsim 2$. As we can see from Fig. \ref{fig:comparebiasPS}, at
lower redshifts, or equivalently lower $M_0/M_\star$, our predictions for the
quasar power spectrum should be more uncertain.

One might imagine modifying the above model by allowing mergers to take place
preferentially in massive parents, and therefore boosting the predicted bias.
In model B, we adopt a more general procedure of matching the observed quasar
abundance by $L (d\Phi/dL) = M_1 \int_{M_1}^\infty dM [dN(M, z_0)/dM] [d{\cal
N}(M_1, z_1 | M, z_0) / dM_1]$ $(t_Q/t_{\rm Hub}) f(M_1, M) $ where
$(t_Q/t_{\rm Hub}) f(M_1, M)$ is the probability that an $M_1$ sub--halo
residing within a parent halo of $M_0$ harbors an active quasar of luminosity
$L$.  It is conceivable that $f$ increases with the parent mass $M_0$ -- a more
massive parent might encourage more quasar activity by having a higher fraction
of mergers or collisions.  The following heuristic argument shows that one
might expect $[d{\cal N}(M_1, z_1 | M, z_0) / dM_1] f(M_1, M)$ to scale
approximately as $\sim M^{4/3}$. Let $N_h$ be the number of sub--halos inside a
parent halo of mass $M$. The rate of collisions is given by $N_h^2 v_h \sigma_h
/R^3$ where $v_h$ is the velocity of the sub--halos, $\sigma_h$ is their
cross-section and $R^3$ is the volume of the parent halo. Using $N_h \propto M$
(which can be obtained from eq. \ref{eq:condPS} in the large $M$ limit), $v_h
\propto \sqrt{M/R}$ (virial theorem) and $R^3 \propto M$ (fixed overdensity of
$\sim 200$ at the redshift of formation), the rate of collisions scales with
the parent mass as $M^{4/3}$, if one ignores the possibility that $\sigma_h$
might depend on the parent mass as well. A similar scaling of $M^{1.3}$ has
been observed in simulations of the star-burst model for Lyman-break objects
(Kolatt et al. 1999, Weschler et al. 1999).

To model the enhanced rate of collisions inside massive parent halos, we can
simply modify model A by using $f(M_1, M) = (M/M_1)^{1/3}$.  The effective
bias is given by

\begin{eqnarray}
&& b_{\rm eff}^B (L,z_1) =  \\ \nonumber
&& \left[\int_{M_1}^\infty dM {dN(M, z_0) \over dM} {d{\cal N}
(M_1, z_1 | M, z_0) \over dM_1} \left({M\over M_1}
\right)^{1\over 3}  \right]^{-1} \\ \nonumber &&
\times\int_{M_1}^\infty dM {dN(M, z_0) \over dM} {d{\cal N}
(M_1, z_1 | M, z_0) \over dM_1} \left({M \over M_1}\right)^{1\over 3} 
b(M,z_0).
\label{eq:beffB}
\end{eqnarray}

We find that the above prescription does not significantly alter our
conclusions following from model A: the $M^{1/3}$ enhancement of the activation
rate inside massive parent halos turns out to be relatively shallow, and
translate to a small effect in the bias.  Finally, we note that $z_1$ above
could in principle depend on $M_0$ and $M_1$, a possibility that would require
further modeling and is not pursued here.

\vspace{\baselineskip} {\em Galaxies without BH's.} Another possibility that
could modify our picture is that only a fraction $f<1$ of the halos harbor
BH's; the duty cycle could then reflect this fraction, rather than the lifetime
of quasars.  Although there is evidence (e.g. Magorrian et al. 1998) that most
{\it nearby} galaxies harbor a central BH, this is not necessarily the case at
redshifts $z=2-3$: the fraction $f$ of galaxies hosting BH's at $z=2-3$ could,
in principle have merged with the fraction $1-f$ of galaxies without BH's,
satisfying the local constraint.

Using the extended Press--Schechter formalism (Lacey \& Cole 1993), one can
compute the rate of mergers between halos of various masses.  On galaxy--mass
scales, the merger rates at $z=2-3$ are comparable to the reciprocal of the
Hubble time (cf. Fig. 5 in Haiman \& Menou 2000), implying that a typical
galaxy did not go through numerous major mergers between $z=2-3$ and $z=0$,
i.e. that the fraction $f$ cannot be significantly less than unity at $z=2-3$.
A more detailed consideration of this issue is beyond the scope of this paper;
we simply note that the lifetimes derived here scale approximately as $1/f$,
where $f$ is likely of order unity.

\vspace{\baselineskip} {\em Larger scatter in $L/M$.}  The scatter $\sigma$ we
assumed around the mean relation between quasar luminosity and halo mass is
motivated by the scatter found empirically for the $M_{\rm bh}-M_{\rm bulge}$
relation (Magorrian et al. 1998).  It is interesting to consider the
sensitivity of our conclusions to an increased $\sigma$.  In general, scatter
raises the number of quasars predicted by our models, by an amount that depends
on the slope of the underlying mass function $dN/dM$. As a result, increasing
$\sigma$ raises, and flattens the predicted LF.  We find that an increase of
$\sigma$ from 0.5 to 1 (an additional order of magnitude of scatter) can be
compensated by a steeper $\bar L_M$ relation, typically replacing $\alpha$ with
$\approx \alpha-0.5$.  As a result of the increase in $\sigma$, quasars with a
fixed $L$ are, on average, associated with larger, and more highly biased
halos.

Nevertheless, we find that the mean bias $\bar b$ of all sources above a fixed
flux (cf. eq.~\ref{eq:bave}), and therefore the correlation length $r_0$, is
unchanged by the increased scatter (at the level of $\sim 3\%$).  The reason
for the insensitivity of $r_0$ to the amplitude of the scatter can be
understood as follows. The mean bias $\bar b$ of all sources with $L>L_{\rm
min}$ is dominated by the bias $b(L)$ of sources near the threshold $L_{\rm
min}$.  The latter is obtained by averaging $b(M)$ over halos of different
masses (cf. eq.~\ref{eq:biasL}), and it is dominated by the bias of the
smallest halos within the width of the scatter, i.e. of halos with mass $M_{\rm
min} \approx \bar M / 10^\sigma$, where $\bar M$ defines the mean relation
between $L_{\rm min}$ and halo mass i.e. $L_{\rm min}$ = $\bar L(\bar M)$, and
$\sigma$ quantifies the scatter (see eq. \ref{eq:scatter}).  As mentioned
before, increasing the scatter makes the luminosity function flatter, which
means to match the observed abundance of halos at a fixed luminosity $L_{\rm
min}$, one has to choose a higher $\bar M$. In other words, $\bar M$ scales up
with the scatter, and it turns out to scale up approximately as $10^\sigma$,
making $M_{\rm min}$ and hence the effective bias roughly independent of
scatter.

We note that the relation between quasar luminosity and halo mass can, in
principle be derived from observations, by measuring $M_{\rm halo}$ for the
hosts of quasars (e.g. by weak lensing, or by finding test particles around
quasars, such as nearby satellite galaxies).

\vspace{\baselineskip} {\em Mass and redshift dependent lifetime.} In all of
the above, we have assumed that the quasar lifetime is a single parameter,
independent of the halo mass.  This is not unreasonable if the Eddington time,
the timescale for the growth of black hole mass, is indeed the relevant
time--scale, $4\times10^7$ ($\epsilon$/0.1) yr.  Implicit in such reasoning is
that the active phase of the quasar is coincident with the phase where the
black hole gains most of its mass. This is not the only possibility; see
Haehnelt et al. (1999) for more discussions. One can attempt to explore how
$t_Q$ depends on halo mass by applying our method to quasars grouped into
different absolute luminosity ranges, but the intrinsic scatter in the
mass-luminosity relation should be kept in mind.  We emphasize, however, since
we fit the luminosity function and clustering data at the same redshift, there
is no need within our formalism to assume a redshift independent lifetime. In
fact, performing our exercise as a function of redshift could give interesting
constraints on how $t_Q$ evolves with redshift.

\section{Conclusions}
\label{section:conclusions}

In this paper, we have modeled the quasar luminosity function in detail in the
optical and X--ray bands using the Press--Schechter formalism.  The lifetime of
quasars $t_Q$ enters into our analysis through the duty--cycle of quasars, and
we find that matching the observed quasar LF to dark matter halos yields the
constraint $10^6 \lsim t_Q \lsim 10^8$ yr: smaller lifetimes would imply overly
massive BH's, while longer lifetimes would necessitate overly massive halos.
This range reassuringly brackets the Eddington timescale of $4\times10^7$
($\epsilon$/0.1) yr.

The main conclusion of this paper is that the lifetime (and hence $\epsilon$,
if the Eddington time is the relevant timescale for quasar activity) can be
further constrained within this range using the clustering of quasars: for
quasars with a fixed luminosity, longer $t_Q$ implies larger host halo masses,
and higher bias.  We find that as a result, the correlation length $r_0$ varies
strongly with the assumed lifetime.  Preliminary data from the 2dF survey
already sets mild constraints on the lifetime.  Depending on the assumed
cosmology, we find $t_Q=10^{7.7\pm 0.8}$ yr ($\Lambda$CDM) or $t_Q=10^{8\pm
0.8}$ yr (OCDM) to within $1\sigma$ statistical uncertainty. These values are
also found to satisfy upper limits on the auto--correlation function of the
soft X--ray background.

Forthcoming large quasar samples from SDSS or the complete 2dF survey are
ideally suited for a study of quasar clustering, and they can, in principle
constrain the quasar lifetime to high accuracy, with small statistical
errors. We expect the modeling of the quasar--halo relation, as well as the
possible presence of obscured quasars, to be the dominant sources of systematic
uncertainty.  Not discussed in depth in this paper is the possibility of using
higher moments (such as skewness), which our models also make definite
predictions for, and will be considered in a future publication. Remarkably,
our best determination of the lifetime of quasars might come from the
statistics of high--redshift quasars, rather than the study of individual
objects.

\acknowledgments 

Near the completion of this work, we became aware of a similar, independent
study by P. Martini \& D. Weinberg. We thank M. Haehnelt, K. Menou,
E. Quataert, U-L. Pen., U. Seljak, R. Sheth and D. Spergel for useful
discussions, T. Miyaji for his advice on the X--ray luminosity function, and
T. Shanks and S. Croom for discussions on the 2dF survey. Support for this work
was provided by the DOE and the NASA grant NAG 5-7092 at Fermilab, by the NSF
grant PHY-9513835, by the Taplin Fellowship to LH and by NASA through the
Hubble Fellowship grant HF-01119.01-99A to ZH, awarded by the Space Telescope
Science Institute, which is operated by the Association of Universities for
Research in Astronomy, Inc., for NASA under contract NAS 5-26555.

\section*{Appendix}

In this Appendix, we describe our models for the luminosity function (LF) of
quasars, based on associating quasar BHs with dark halos.  Our main assumption
is that there is, on average, a direct monotonic relation between halo mass and
quasar light.  Our treatment is similar to previous works (Haiman \& Loeb 1998;
Haehnelt et al. 1998), but differs in some of the details.  We adopt the
parameterization of the observational LF in the optical B band, given in the
redshift range $0<z\lsim 4$ by Pei (1995).  We assume the background cosmology
to be either flat ($\Lambda$CDM) with $(\Omega_\Lambda,\Omega_{\rm
m},h,\sigma_{\rm 8h^{-1}},n)=(0.7,0.3,0.65,1.0,1.0)$ or open (OCDM) with
$(\Omega_\Lambda,\Omega_{\rm m},h,\sigma_{\rm
8h^{-1}},n)=(0,0.3,0.65,0.82,1.3)$.  The LF quoted by Pei (1995) is scaled
appropriately with cosmology by keeping $(d\phi/dL) dV d_{\rm L}^2$=const,
where $dV$ is the volume element, and $d_{\rm L}$ is the luminosity distance),
so that $(d\phi/dL)dL$ is the comoving abundance in ${\rm Mpc^{-3}}$ of quasars
with B--band luminosity $L$ (in solar units $L_{B,\odot}$).  The comoving
abundance $dN/dM(M,z)$ of dark halos is assumed to follow the Press--Schechter
(1974) formalism. We assume that each halo harbors a single quasar that turns
on when the halo forms, i.e. triggered by merger (e.g. Percival \& Miller
1999), and shines for a fixed lifetime $t_Q$ (relaxing these assumptions is
discussed below in section~\ref{section:discussion}).  The duty--cycle $f_{\rm
on}$ of halos with mass $M$ at redshift $z$, is then given by the fraction of
these halos younger than $t_Q$. The distribution of ages $dp_a/dt(M,z,t)$ for
halos of mass $M$ existing at redshift $z$ is obtained using the extended
Press-Schechter formalism, which assumes that the halo formed at the epoch when
it acquired half of its present mass (Lacey \& Cole 1993).  The duty--cycle,
which is the probability that a dark matter halo of a given mass harbors an
active quasar, is simply

\beq
f_{\rm on}(M,z)=\int_0^{t_Q} dt \frac{dp_a}{dt}(M,z,t).
\label{eq:duty}
\eeq

A model in which the quasar turns on/off more gradually (as expected if the
mass of the BH grows significantly during the luminous quasar phase) is
equivalent to one having additional scatter in the ratio $L/M$, which is
discussed in \S~\ref{section:discussion}.  We next relate the quasar luminosity
to the mass of its host halo.  We define $dp/dL(M,L,z)$ to be the probability
that a halo of mass $M$ at redshift $z$ hosts a quasar with luminosity $L$, and
express this quantity as

\beq
\frac{dp}{dL}(L,M,z) = \frac{dg}{dL}(L,\bar L_{M,z}) f_{\rm on}(M,z).
\label{eq:pdiff}
\eeq

Here $dg/dL(L,\bar L_{M,z})$ is the probability distribution of luminosities
associated with the subset of halos of mass $M$ harboring a live quasar
(normalized to $\int_0^\infty dL dg/dL=1$), and $\bar L_{M,z}$ is the mean
quasar luminosity for these halos.  In the limit of a perfect intrinsic
correlation, we would have $dg/dL(L,\bar L_{M,z})=\delta(L-\bar L_{M,z})$; more
realistically, this correlation will have non--negligible scatter.  Lacking an
a--priori theory for this scatter, we here simply assume that it follows the
same functional form as the scatter found empirically for the $M_{\rm
bh}-M_{\rm bulge}$ relation (Magorrian et al. 1998), and we set

\beq
\frac{dg}{dL}(L,\bar L_M) \propto \exp(-(\log L/\bar L_{M,z})^2/2\sigma^2).
\label{eq:scatter}
\eeq

For reference, we note that the empirical scatter between $M_{\rm bh}$ and
$M_{\rm bulge}$ gives $\sigma\sim0.5$, it is not yet clear, however, what
fraction of this scatter is intrinsic vs. instrumental (van der Marel
1999). One might expect the scatter in the $L$--$M_{\rm halo}$ relation not to
be significantly larger, since (i) for a sufficiently high fueling rate, the
luminosity $L$ corresponding to $M_{\rm bh}$ is likely to always be near the
Eddington limit, and (ii) at least for disk galaxies, the bulge luminosity
correlates well with the total luminosity $L_{\rm tot}$ ($\sigma\sim0.5$, see
e.g. Andredakis \& Sanders 1994); $L_{\rm tot}$ is tightly correlated with the
velocity dispersion $\sigma_v$ through the Tully-Fisher relation
(e.g. Raychaudhury et al. 1997) as is the total halo mass to $\sigma_v$
(Eisenstein \& Loeb 1996).  Nevertheless, in \S~\ref{section:discussion} below
we will investigate the consequences of an increased scatter.  We note that an
extension of the models presented here, by following the merger histories of
halos and their BH's can, in principle, be used to estimate the scatter in
$L/M_{\rm halo}$.  Cattaneao, Haehnelt \& Rees (1998) have used this approach
to fit the observed relation $M_{\rm bh}/M_{\rm bulge}$, including its scatter.

Under the above assumptions, the cumulative probability that a halo of mass $M$
hosts a quasar with luminosity equal to or greater than $L$ is given by

\beq
p(L,M,z)=f_{\rm on}(M,z) \int_L^\infty dL \frac{dg}{dL}(L,\bar L_{M,z}),
\label{eq:pcum}
\eeq

and matching the observed cumulative quasar LF requires

\beq
\int_L^\infty dL \frac{d\phi}{dL}(L,z)
 = \int_0^\infty dM  \frac{dN}{dM}(M,z) p(L,M,z),
\label{eq:matchcum}
\eeq

or alternatively, matching the differential LF gives

\beq
\frac{d\phi}{dL}(L,z) = \int_0^\infty dM  \frac{dN}{dM}(M,z) 
\frac{dg}{dL}(L,\bar L_{M,z}) f_{\rm on}(M,z).
\label{eq:matchdiff}
\eeq

\begin{figure}[t]
\includegraphics{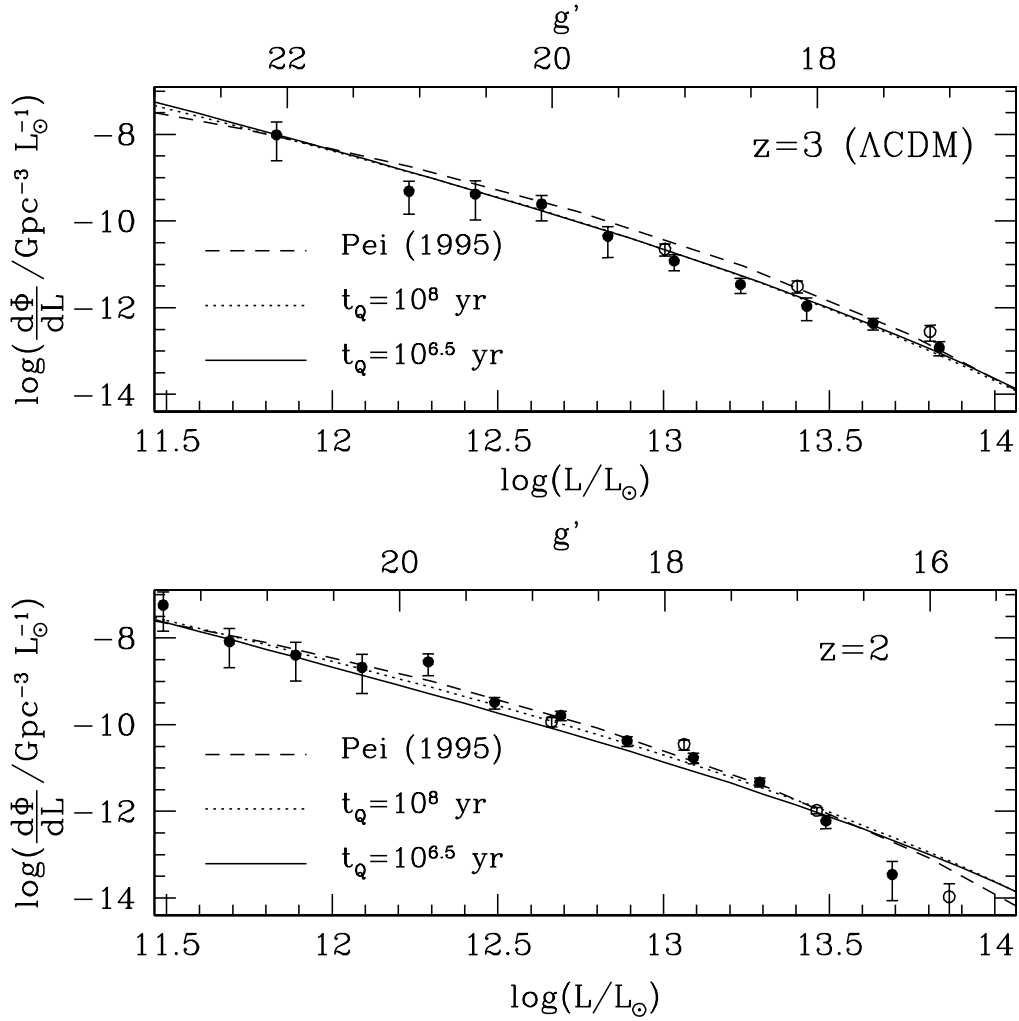}
\vspace*{4.5in}
\caption{Fits to the quasar luminosity function at redshifts $z=2$ and 3 in our
models, with two different quasar lifetimes $t_Q=10^{6.5}$ (solid curves)
and $t_Q=10^8$ yr (dotted curves).  Also shown are the data, and fitting
function (dashed curves) for the LF from Pei (1995).  The quality of our fits
at different redshifts or in the OCDM model are similar.  The upper labels show
the corresponding apparent magnitudes in the SDSS $g^\prime$ band, assuming
that the intrinsic quasar spectrum is the same as the mean spectrum in the
Elvis et al. (1994) quasar sample.  }
\label{fig:LFfits}
\end{figure}

\begin{figure}[t]
\includegraphics{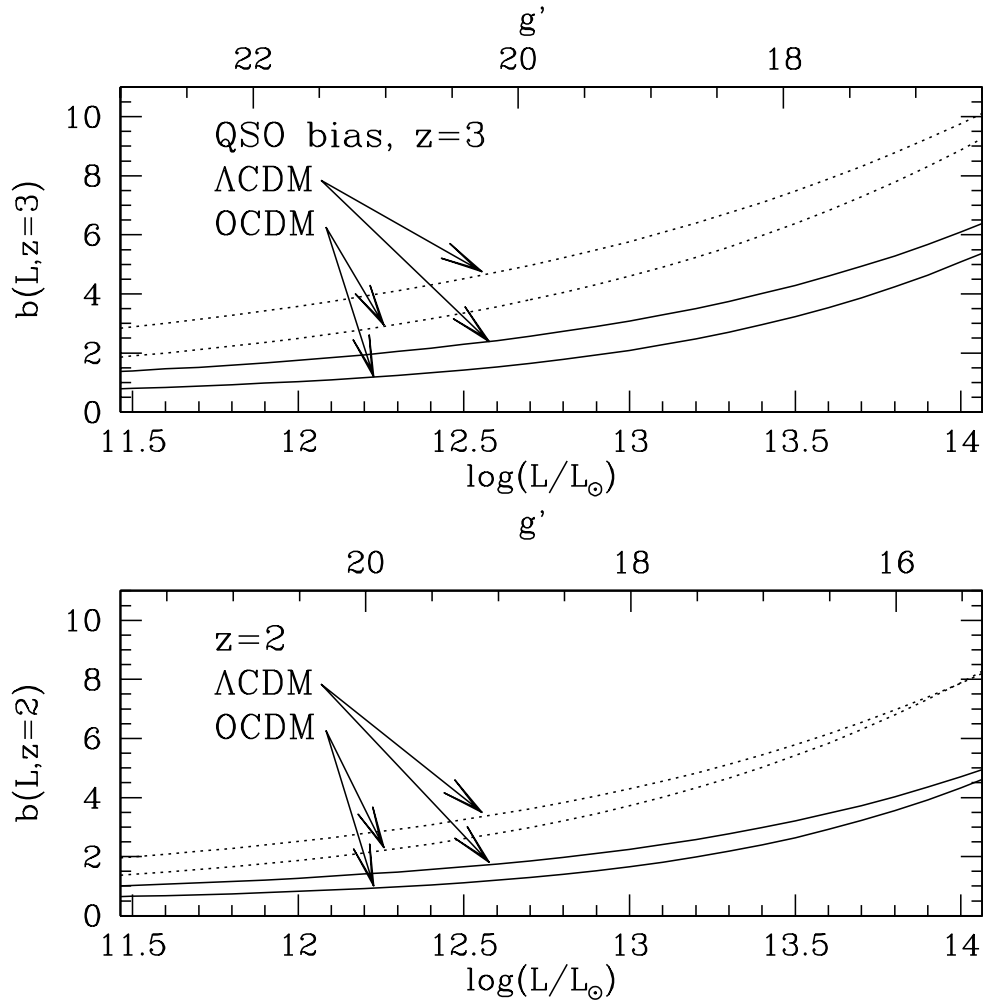}
\vspace*{4.5in}
\caption{Bias parameter $b(L,z)$ of quasars as a function of their intrinsic
B--band luminosity (lower labels) or apparent SDSS $g^\prime$ magnitude (upper
labels), in the models with two different quasar lifetimes $t_Q=10^{6.5}$
(solid curves) and $t_Q=10^8$ yr (dotted curves) as shown in
Figure~\ref{fig:LFfits}.  For comparison, the bias parameter is also shown in
the OCDM model.  Quasars are more highly biased in the long lifetime models,
and in $\Lambda$CDM.}
\label{fig:bias}
\end{figure}

\begin{figure}[t]
\includegraphics{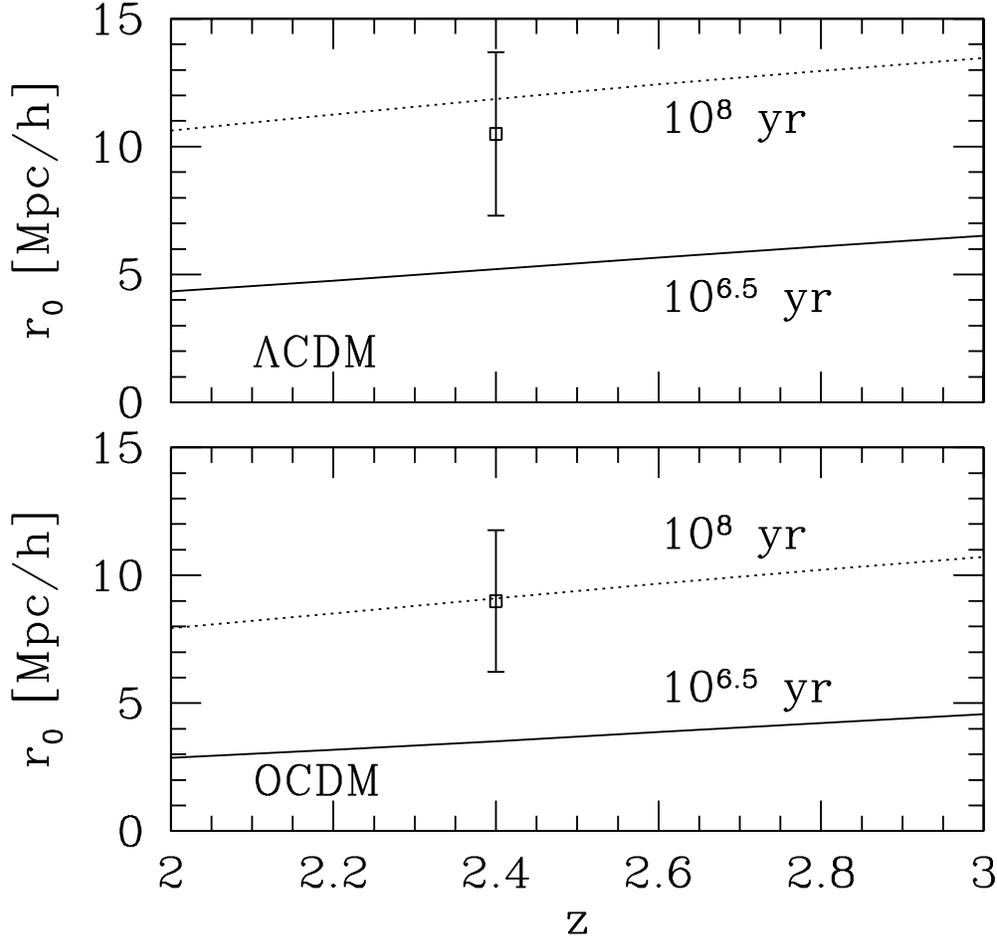}
\vspace*{4.5in}
\caption{Correlation length $r_0$ in our models with two different quasar
lifetimes $t_Q=10^{6.5}$ (solid curves) and $t_Q=10^8$ yr (dotted curves).  An
apparent magnitude cut of $B<20.85$ was used, corresponding to the limits of
the 2dF survey (Croom et al. 1999). The open square shows a preliminary result
from 2dF. The upper panel shows our results in the $\Lambda$CDM model, and the
lower panel in OCDM.}
\label{fig:r0}
\end{figure}

\begin{figure}[t]
\includegraphics{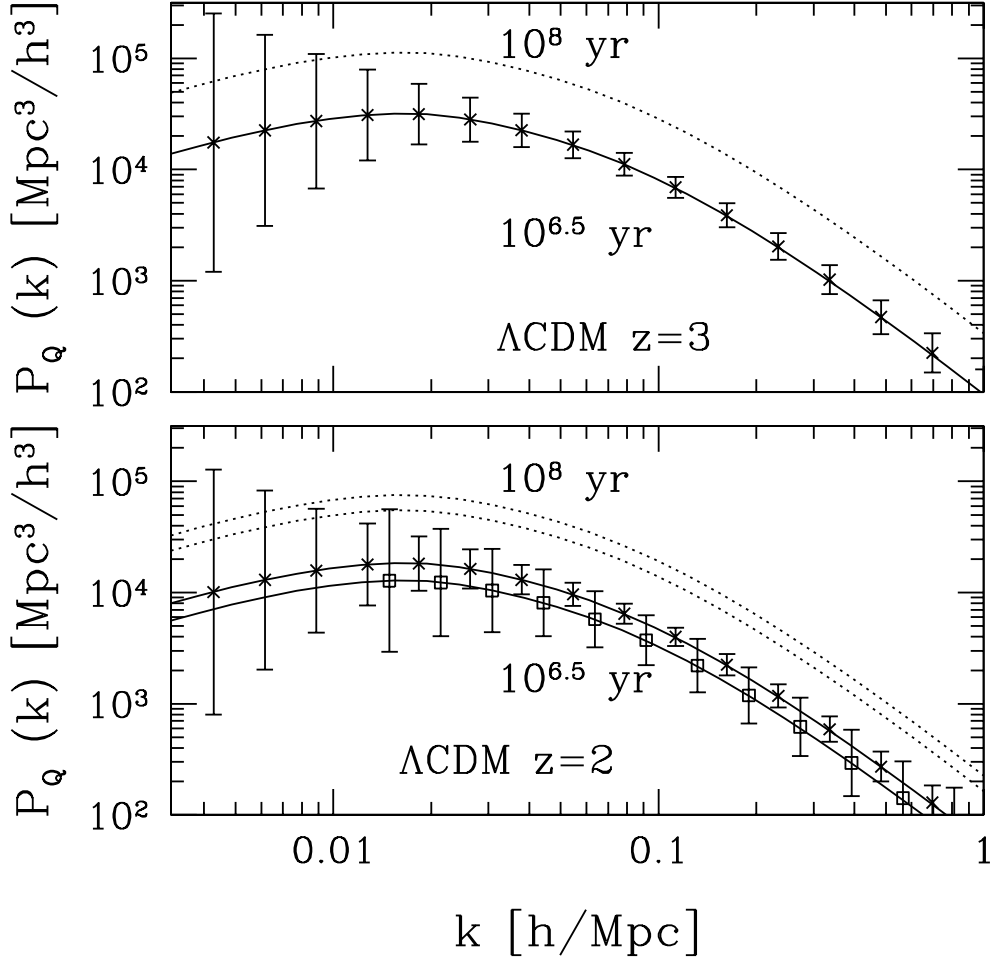}
\vspace*{4.5in}
\caption{Three--dimensional power spectrum $P_Q(k)$ of quasars in $\Lambda$CDM
at two different redshifts near the peak of the comoving quasar abundance.
Results are shown for quasars with $B\leq 20.4$ (or $g^\prime\lsim 19$), in the
long (dotted curves) and short lifetime (solid curves) models, together with
the expected $1\sigma$ error bars from SDSS (crosses) with an assumed area of
$\pi$ steradians.  The slightly lower curves in the lower panel refer to 2dF
(open squares), with a magnitude cut of $B=20.5$, and show the expected error
bars from an assumed area of 0.23 steradians.}
\label{fig:sdsslcdm}
\end{figure}

\begin{figure}[t]
\includegraphics{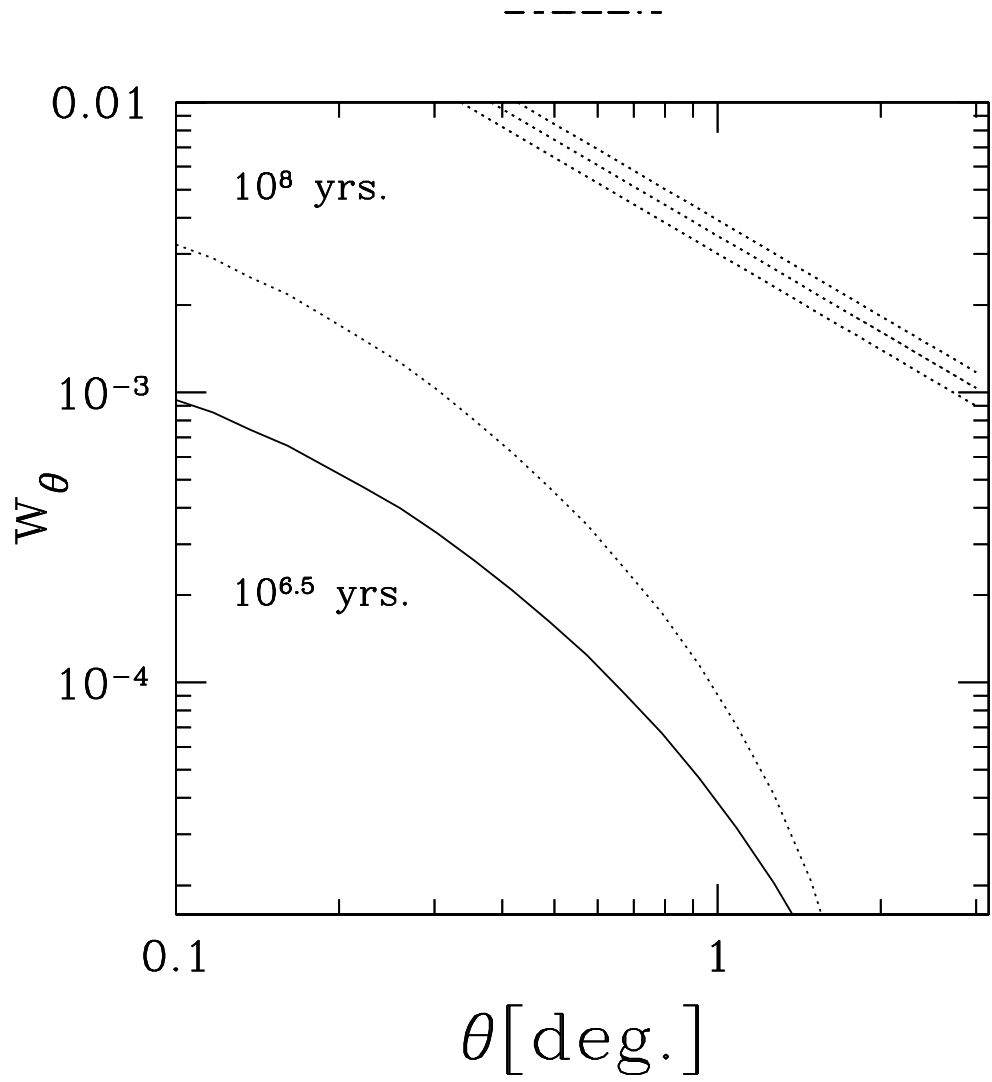}
\vspace*{4.5in}
\caption{Angular correlation function of the total XRB, $w_\theta$ at $E = 1$
keV. The dashed lines in the upper right corner indicate $w_\theta$ at $E =
1.15$ keV with quoted $\pm 1\sigma$ uncertainties as measured from the ROSAT
All Sky Survey (Soltan et al. 1999), which can be considered an upper limit.}
\label{fig:xrb}
\end{figure}

\begin{figure}[t]
\includegraphics{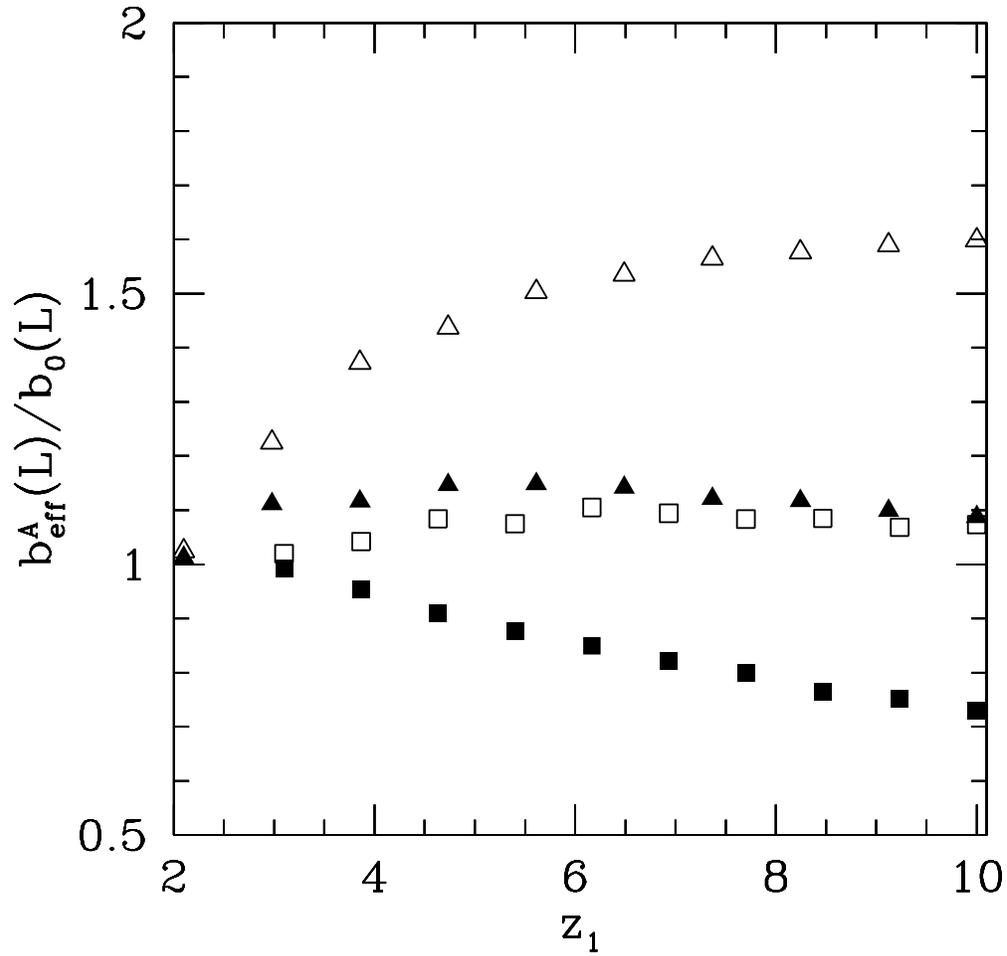}
\vspace*{4.5in}
\caption{The ratio of the effective bias $b_{\rm eff}^A (L)$ (in a model
allowing multiple quasars per halo) to the original bias $b_0 (L)$ (for one
quasar per halo) as a function of $z_1$ at fixed luminosity $L$, where $z_1$ is
the redshift when sub--halos are identified.  The squares (open for $M_0 =
10^{12.5} M_\odot$ and solid for $M_0 = 10^{13.5} M_\odot$) show this relation
for quasars at $z_0 = 3$, whereas the triangles are for $z_0 = 2$ (open for
$M_0 = 10^{12} M_\odot$ and solid for $M_0 = 10^{13} M_\odot$). The model is
$\Lambda$CDM.  See text for discussions.}
\label{fig:comparebiasPS}
\end{figure}

Equation \ref{eq:matchcum} or \ref{eq:matchdiff}, together with equations
\ref{eq:duty}, \ref{eq:scatter}, and \ref{eq:pcum} implicitly determines the
function $\bar L_{M,z}$, once the quasar lifetime $t_Q$ and magnitude of
scatter $\sigma$ are specified.  In general, these equations would need to be
solved iteratively. In practice, we have found that sufficiently accurate
solutions (given the error bars on the observational LF in Pei 1995;
cf. Fig.~\ref{fig:LFfits}) can be found by using the simple power--law ansatz:

\beq
\bar L_{M,z}= x_0(z) M_{\rm halo} 
              \left(\frac{M_{\rm halo}}{M_0}\right)^{\alpha(z)},
\label{eq:params}
\eeq

where the coefficients $x_0(z)$ and $\alpha(z)$ depend on $t_Q$, $\sigma$, and
the underlying cosmology (Haiman \& Loeb 1998; Haehnelt et al. 1998).  In
summary, assuming a fixed scatter $\sigma$, our model has only one free
parameter, the quasar lifetime $t_Q$.

We emphasize that our parameterization in equation \ref{eq:params} is purely
phenomenological -- it gives us a convenient way to relate the quasar
luminosity to the host halo mass ($\bar L_{M,z}$).  In reality, the quasar
luminosity likely depends on the details of its immediate physical environment
(e.g. gas supply, magnetic fields, angular momentum distribution, etc.), in
addition to the halo mass.  Our description includes these possibilities only
in allowing a non--negligible scatter around the mean relation $\bar L_{M,z}$.
The rationale behind this choice is that the average properties of the physical
environment should ultimately be governed by the halo mass (or circular
velocity), as expected within the picture of structure formation via
hierarchical clustering.

A useful check on the physical implications of equation \ref{eq:params} is
obtained by assuming that the luminosity $\bar L_{M,z}$ is produced by a BH of
mass $M_{\rm bh}$, shining at the Eddington limit $L_{\rm Edd}=(4\pi G \mu m_p
c/ \sigma_{T}) M_{\rm bh}$. In the mean spectrum of a sample of quasars with
detections from radio to X--ray bands (Elvis et al. 1994), $\approx 7\%$ of the
bolometric luminosity is emitted in the rest--frame $B$ band, resulting in
$L=0.07 L_{\rm Edd} = 5\times 10^3 {\rm L_{B,\odot} (M_{\rm bh}/M_\odot})$.
Equation \ref{eq:params} then translates into a relation between the mass of a
BH and its host halo,

\beq
\bar M_{\rm bh} = 10^{-3.7} x_0(z) M_{\rm halo} 
              \left(\frac{M_{\rm halo}}{M_0}\right)^{\alpha(z)}.
\label{eq:params-bh}
\eeq

\begin{figure}[t]
\includegraphics{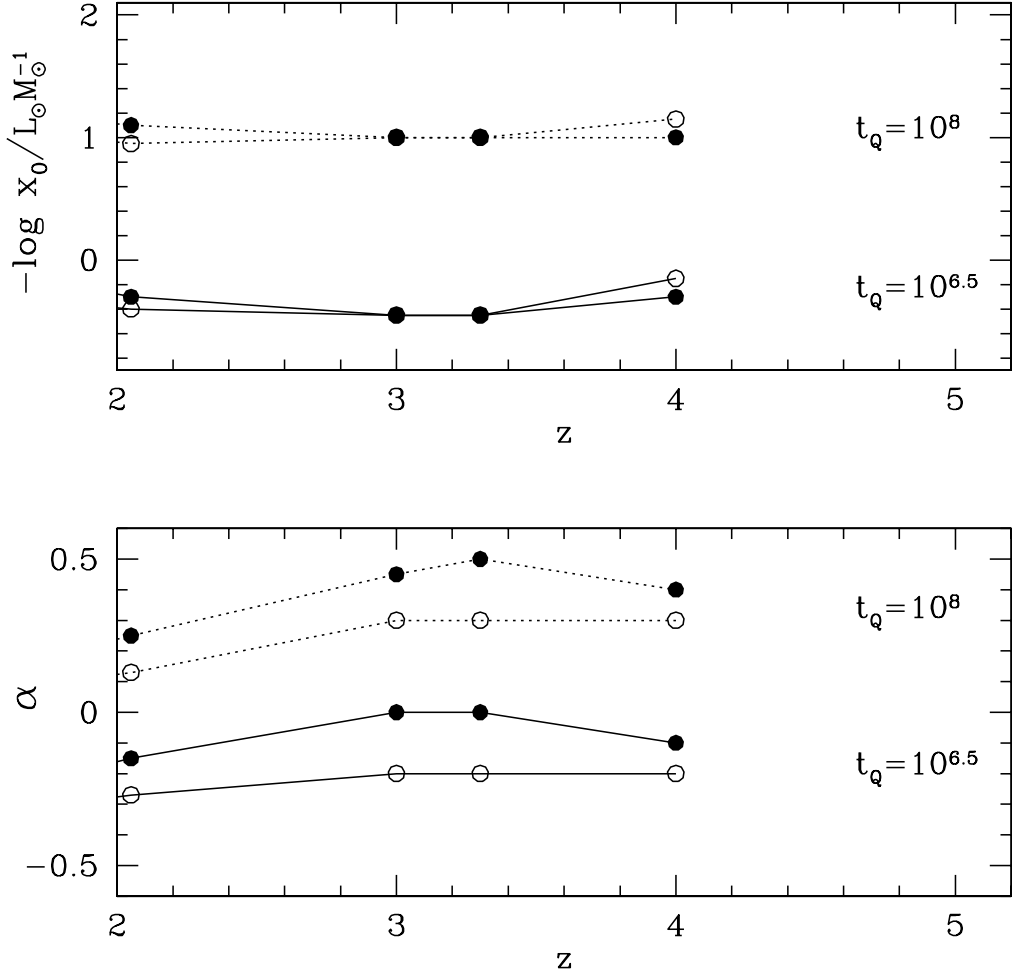}
\vspace*{4.5in}
\caption{Fitting parameters $\alpha(z)$ and $x_0(z)$ for the mean relation
between B--band quasar luminosity $L_B$ and halo mass $M_{\rm halo}$ given in
equation \ref{eq:params}, for two different quasar lifetimes $t_Q=10^{6.5}$
(solid curves) and $t_Q=10^8$ yr (dotted curves).  The filled dots correspond
to a $\Lambda$CDM and the open dots to an OCDM cosmology.  In all cases, we
assumed a scatter with $\sigma=0.5$ (cf. eq.~\ref{eq:scatter}) around the mean
$L-M_{\rm halo}$ relation.}
\label{fig:params}
\end{figure}

As an example, Haehnelt et al. (1999) argue that the central BH mass is
determined by a radiative feedback from the central BH that would unbind the
disk in a dynamical time. Their derived scaling corresponds to $\alpha=2/3$ and
$x_0\propto (1+z)^{5/2}$, not far from what we find for the long--lifetime case
(cf. Figure~\ref{fig:params} and discussion below).

In Figure~\ref{fig:params}, we show the values of the parameters $x_0(z)$ and
$\alpha(z)$ obtained in our models when two different quasar lifetimes are
assumed, $t_Q=10^{6.5}$ (solid curves) and $t_Q=10^8$ yr (dotted curves).  The
filled dots show the parameters in $\Lambda$CDM, and the empty dots in the OCDM
cosmology.  We have set the arbitrary constant $M_0=10^{12}~{\rm M_\odot}$ in
both cases.  Note that $t_Q$ determines both $\alpha$ and $x_0$, and therefore
the values of $\alpha$ and $x_0$ are correlated.  In general, the fitting
parameters show little evolution in the range $2<z<4$, around the peak of the
quasar LF.  According to equation \ref{eq:params-bh}, the corresponding BH
masses in, e.g. a $10^{12} {\rm M_\odot}$ halo at $2<z<4$ are $M_{\rm
bh}\approx 4\times 10^{-4} M_{\rm halo}= 4\times10^8{\rm M_\odot} $ and $M_{\rm
bh}\approx 2\times 10^{-5} M_{\rm halo}= 2\times10^7{\rm M_\odot}$ in the short
and long lifetime models, respectively.

\begin{figure}[t]
\includegraphics{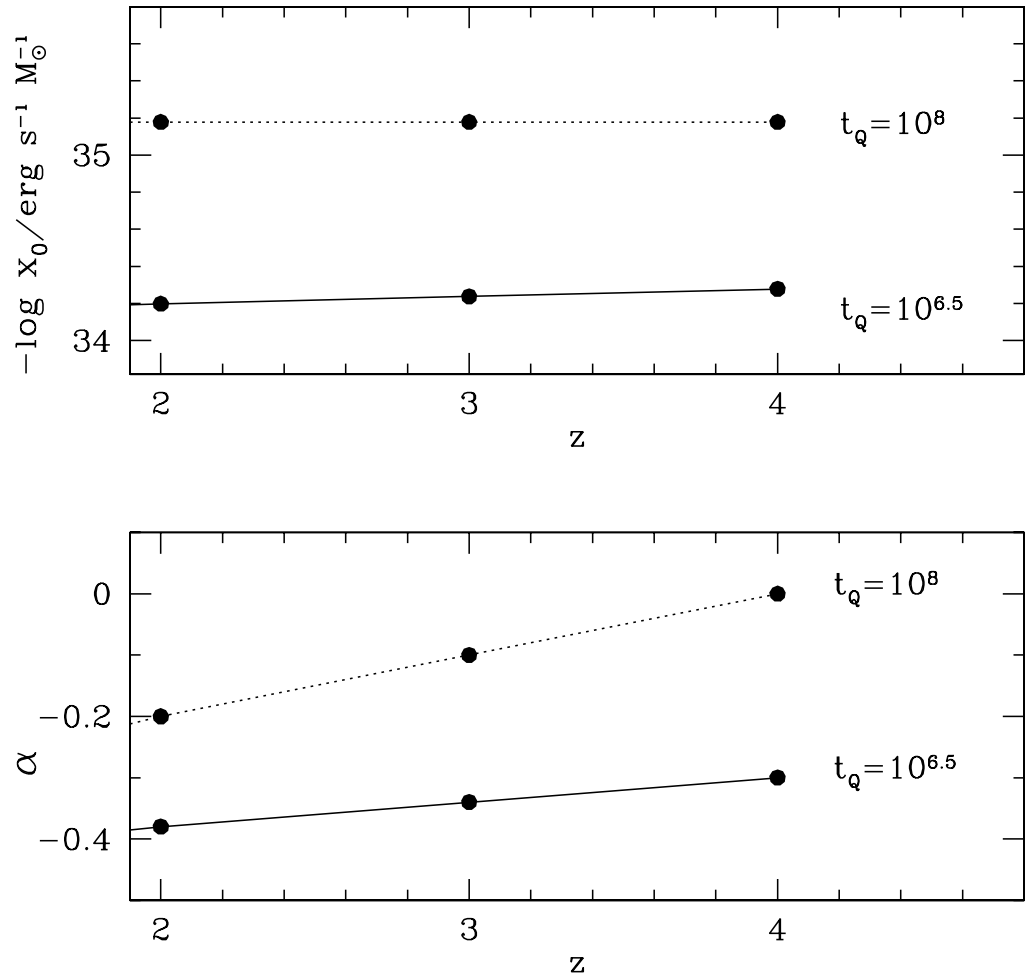}
\vspace*{4.5in}
\caption{Same as Figure~\ref{fig:params}, except the X--ray quasar luminosity
is used at 1keV. Only the $\Lambda$CDM case is shown.}
\label{fig:paramsx}
\end{figure}

The fitting procedure described above can be repeated in the X--ray bands.  We
therefore fit the XRLF using equation~\ref{eq:params} analogously to the
optical case, except $\bar L_{M,z}$ now denotes the X--ray luminosity at 1 keV,
quoted in units of ${\rm erg~s^{-1}}$.  Note that the XRLF in Miyaji et
al. (2000) is quoted a function of luminosity at observed 1 keV, i.e. no
K--correction is applied (alternatively, the XRLF can be interpreted as the
rest--frame luminosity function of sources with an average intrinsic photon
index of 2). Figure~\ref{fig:paramsx} show the resulting fitting parameters
$x_0$ and $\alpha$ in the $\Lambda$CDM cosmology, analogous to those shown in
Figure~\ref{fig:paramsx} for the optical case.  It is apparent that both
parameters have a somewhat behavior different from that in the optical.  This
reflects the fact that the mean quasar spectrum must evolve with redshift, or
at least is black-hole/halo mass dependent: if every quasar had the same
spectrum, or at least a similar X--ray/optical flux ratio, the fitting
parameters derived from the optical and X--ray LF would differ only by a
constant in $x_0$.  For our purpose of deriving clustering, it is sufficient to
treat $x_0$ and $\alpha$ as phenomenological fitting parameters, and we do not
address the physical reason for the apparent spectral evolution (see Haiman \&
Menou for a brief discussion).

It is important to note that the simple power-law ansatz in
equation~\ref{eq:params} with the parameters shown in Figure~\ref{fig:paramsx}
adequately fits only the faint end of the XRLF.  In the optical case, the
entire range of observed luminosities is well matched by our models
(cf. Fig~\ref{fig:LFfits}).  In comparison, the well--fitted range in X--rays
typically extends from the detection threshold to up to 2-3 orders of magnitude
in luminosity (i.e. typically upto $\sim 3\times 10^{45}~{\rm erg~s^{-1}}$),
depending on redshift, and our models underestimate the abundance of still
brighter quasars.  One might then consider searching for a different ansatz to
replace equation~\ref{eq:params} that fits the entire range of the observed
LF. However, we have verified that the rare quasars with these high
luminosities would contribute negligibly both to the clustering signals, or the
XRB investigated here. Therefore, we did not consider further improvements over
equation~\ref{eq:params}, since this would not change our results.

\clearpage
\newpage

\clearpage
\newpage

\clearpage
\newpage

\clearpage
\newpage

\clearpage
\newpage

\newpage

\clearpage
\newpage

\clearpage
\newpage

\end{document}